\documentclass[twocolumn,showpacs,preprintnumbers,amsmath,amssymb]{revtex4}

\usepackage{graphicx}
\usepackage{dcolumn}
\usepackage{bm}

\newcommand{\be}{\begin{equation}}
\newcommand{\ee}{\end{equation}}
\newcommand{\bea}{\begin{eqnarray}}
\newcommand{\eea}{\end{eqnarray}}

\newcommand{\integer}{\relax{\rm I\kern-.18em N}}

\begin{document}

\title{The pentaquark $\Theta^+(1540)$ in the string model}
\author{B.V. Martemyanov$^{1,2}$}
\author{C. Fuchs$^1$}
\author{Amand Faessler$^1$}
\author{M.I. Krivoruchenko$^{1,2}$}
 \affiliation{$^1$ Institut f\"{u}r Theoretische Physik$\mathrm{,}$ 
Universit\"{a}t 
T\"{u}bingen$\mathrm{,}$, 
Auf der Morgenstelle 14$\mathrm{,}$D-72076 T\"{u}bingen, Germany\\
$^2$ Institute for Theoretical and Experimental Physics, 
B. Cheremushkinskaya 25, 117259 Moscow, Russia}
\date{}
\begin{abstract}
We consider the $\Theta^+(1540)$ pentaquark in the string model
that correctly reproduces the linear Regge trajectories for
the case of orbital excitations of light $q\bar q$ mesons and $qqq$ baryons.
Assuming (and arguing in favour of) the diquark-antiquark-diquark
($[ud]\bar s [ud]$) clustering of this orbitally excited object we
found its mass about $290 \rm {MeV}$ above the experimental value
$1540 \rm {MeV}$. In the model considered this discrepancy could be 
attributed to the change of the constituent mass of $\bar s$ antiquark as compared to
that of $s$ quark localized at the string end.  

\end{abstract}
\maketitle

The pentaquark $\Theta^+(1540)$ was initially predicted in chiral soliton
model\cite{1} and then experimentally found by LEPS \cite{2}
and DIANA\cite{3} collaborations. Further it was confirmed by many
other collaborations\cite{4,5,6,7,8,9,10}. More than $150$
theoretical works are devoted to this subject. Since the pentaquark's
 chiral soliton quantum numbers are $J^P = 1/2^+$, in the quark model 
 it could correspond to the first orbital excitation of five quark system 
 $uudd\bar s$ needed to compensate the negative parity of strange
 antiquark. The clustering of quarks in five quark system $uudd\bar s$
 to two spin-isospin singlet diquarks $[ud][ud]$ and strange antiquark $\bar s$
 suggested in \cite{Jaffe} uses this orbital excitation to assure the Bose
 statistics for $[ud]$ diquarks that have antisymmetric color function.
 So, in \cite{Jaffe} the pentaquark is considered consisting from two
 diquarks in relative $p$ -wave and strange antiquark in $s$ -wave with
 respect to the center of mass of two diquarks. It is a first orbital
 excitation of $[ud]\bar s [ud]$ system that has no ground (not orbitally
 excited) state due to Bose statistics. We will consider the prediction
 for the mass of this object in the string model. 

It is far known
(see e.g. the review \cite{molodoiya} and references therein)
 that orbital excitations of light hadrons
i.e. of particles consisting from light $u,~d,~s~$ quarks
are well described by linear Regge trajectories
\be J = \alpha^\prime M^2 +const
\label{regge}
\ee
with $J$ and $M$ being the spin and the mass of the particle,
respectively   and $\alpha^\prime \approx 1 \rm {GeV}^{-2}$
being the slope parameter.

 The linear relation between the particle's
spin and the mass squared is a result of relativistic string
(gluoelectric flux tube)\cite{Nambu}
that is formed between light (i.e. with masses small 
in comparison to string parameter ${\alpha^\prime}^{-1/2}\approx 1 \rm {GeV} $)
quark clusters sitting at the ends of the string. The string tension
$\nu$ and the Regge slope parameter $\alpha^\prime$ are related by the
formula
\be \alpha^\prime = \frac{1}{2\pi\nu}~.
\label{string}
\ee

In the real calculations of the masses of orbitally excited hadrons
\cite{molodoiya} the asymptotic linear relation between the
spin and the mass squared (\ref{regge}) starts practically from
the first orbital excitation. The principle ingredient of the string
model used in \cite{molodoiya} was the spin-orbit coupling  provided
by Thomas precession. The latter is also connected to the Lorentz-scalar
character of the forces responsible for confinement~\cite{EF}. From 
experimental point of view there is no systematic spin-orbit interaction
for low orbital excitations. This could be the result of cancelation
between spin-orbit coupling originating from Lorentz-scalar 
confining forces and that from Coulomb forces~\cite{EF}. If one neglect
the spin-orbit coupling  the spin $J$ and the mass $M$ of the particle
lying on the leading Regge trajectory can be found from the following 
formulae representing the ideal relativistic string with point-like masses
sitting at its ends

\bea
J &=& \sum_{i} \frac{\nu}{2\omega^2}\left( \arcsin {(v_i)} - v_i\sqrt 
 {1 - v_i^2} \right) \nonumber \\
 &+&\sum_{i}\frac{m_i v_i^2}{\omega \sqrt {1 - v_i^2}} +\sum_{i}s_i ~~,
\nonumber \\
M &=& \sum_{i} \frac{\nu}{\omega} \arcsin {(v_i)} + \sum_{i}\frac{m_i }
{\sqrt {1 - v_i^2}} ~~,
\nonumber \\ \\
&&\frac{m_i v_i \omega}{\sqrt {1 - v_i^2}} = {\nu} {\sqrt {1 - v_i^2}}~,
\nonumber
\label{string1}
\eea 
where $\omega$ is the angular velocity of rotation, $v_i,~ m_i,~ s_i$ are
the  velocity, mass and spin of the quark cluster at the end of the string, 
respectively.

The masses of the quark clusters
sitting at the ends of the string   can be fitted by 
 $\omega-f$,~$K^*$,~$\phi$,~ $\Lambda$~- trajectories
(see the fits on Figs.1-4)
\be m_{u(d)} = 330 \rm {MeV},~m_{s} = 420 \rm {MeV},~
m_{[ud]} = 350 \rm {MeV}~.
\label{fit}
\ee

\begin{figure}[!htb]
\vspace{1cm}
\begin{center}
\includegraphics[width=.48\textwidth]{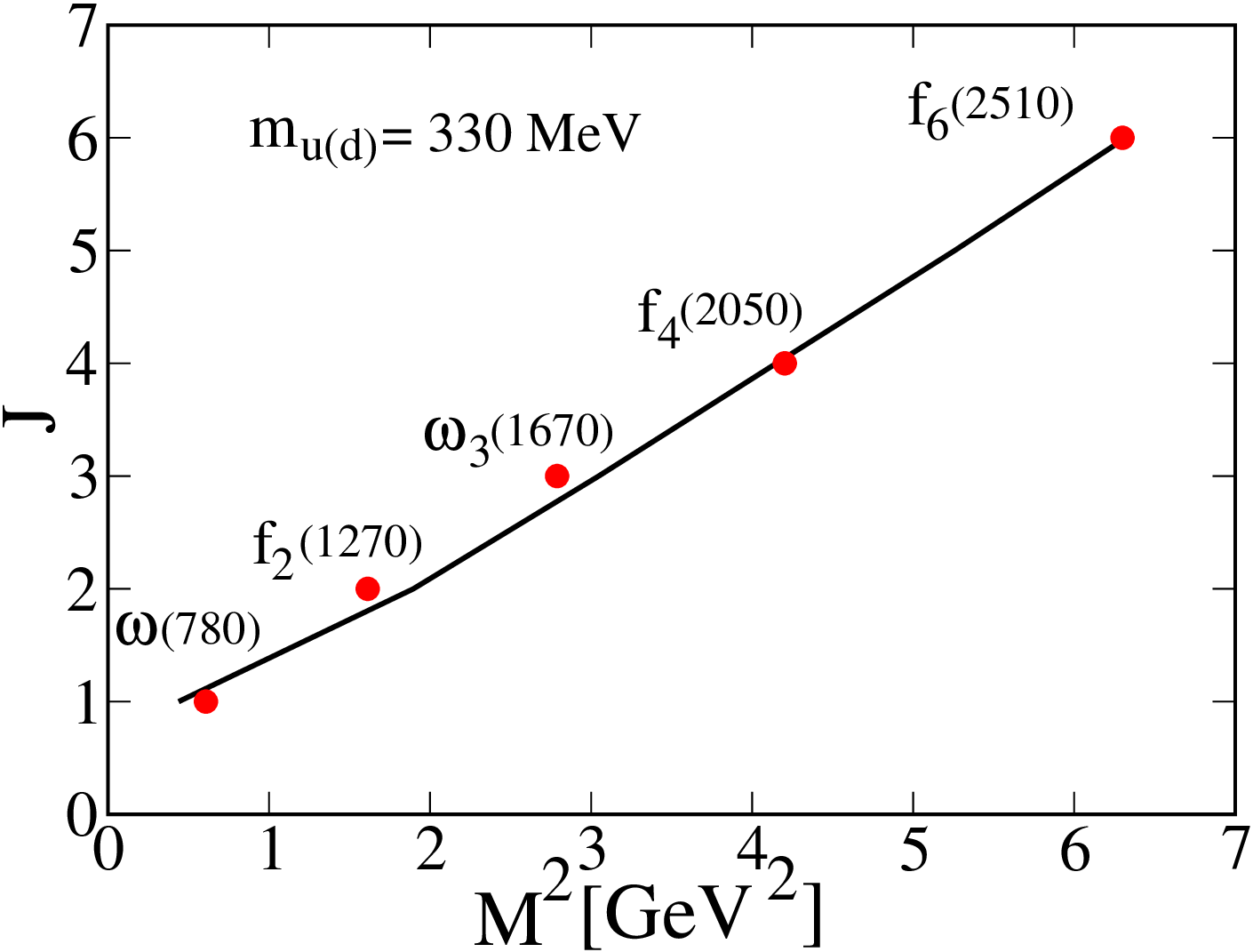}
\caption{The leading $\omega - f$ trajectory fitted by the relativistic
string with nonstrange quarks at the string ends}
\label{reggens}
\end{center}
\end{figure}
\begin{figure}[!htb]
\vspace{1cm}
\begin{center}
\includegraphics[width=.48\textwidth]{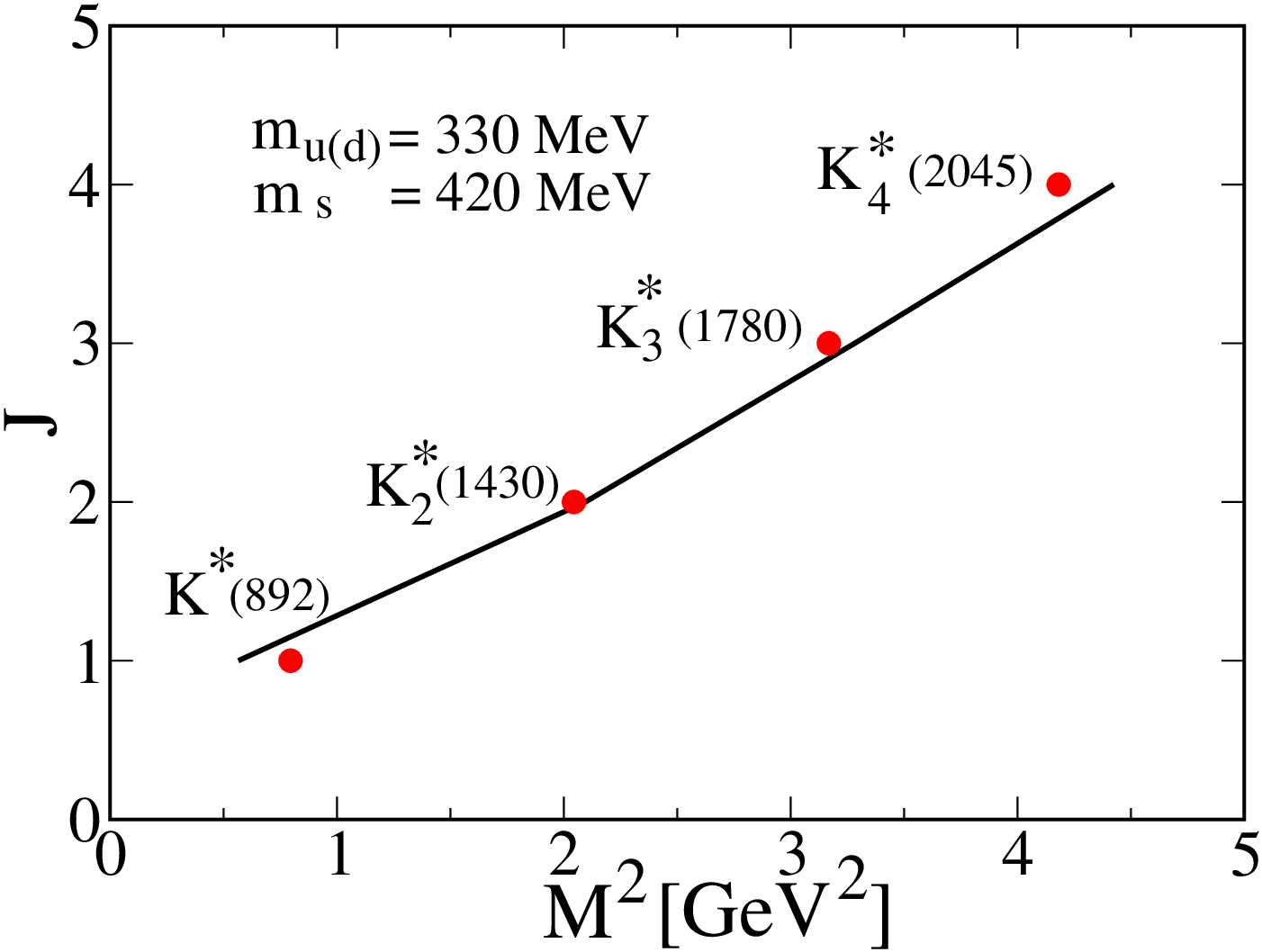}
\caption{The leading $K^*$ trajectory fitted by the  relativistic
string with nonstrange $+$ strange quarks at the string ends}
\label{reggek}
\end{center}
\end{figure}
\begin{figure}[!htb]
\vspace{1cm}
\begin{center}
\includegraphics[width=.48\textwidth]{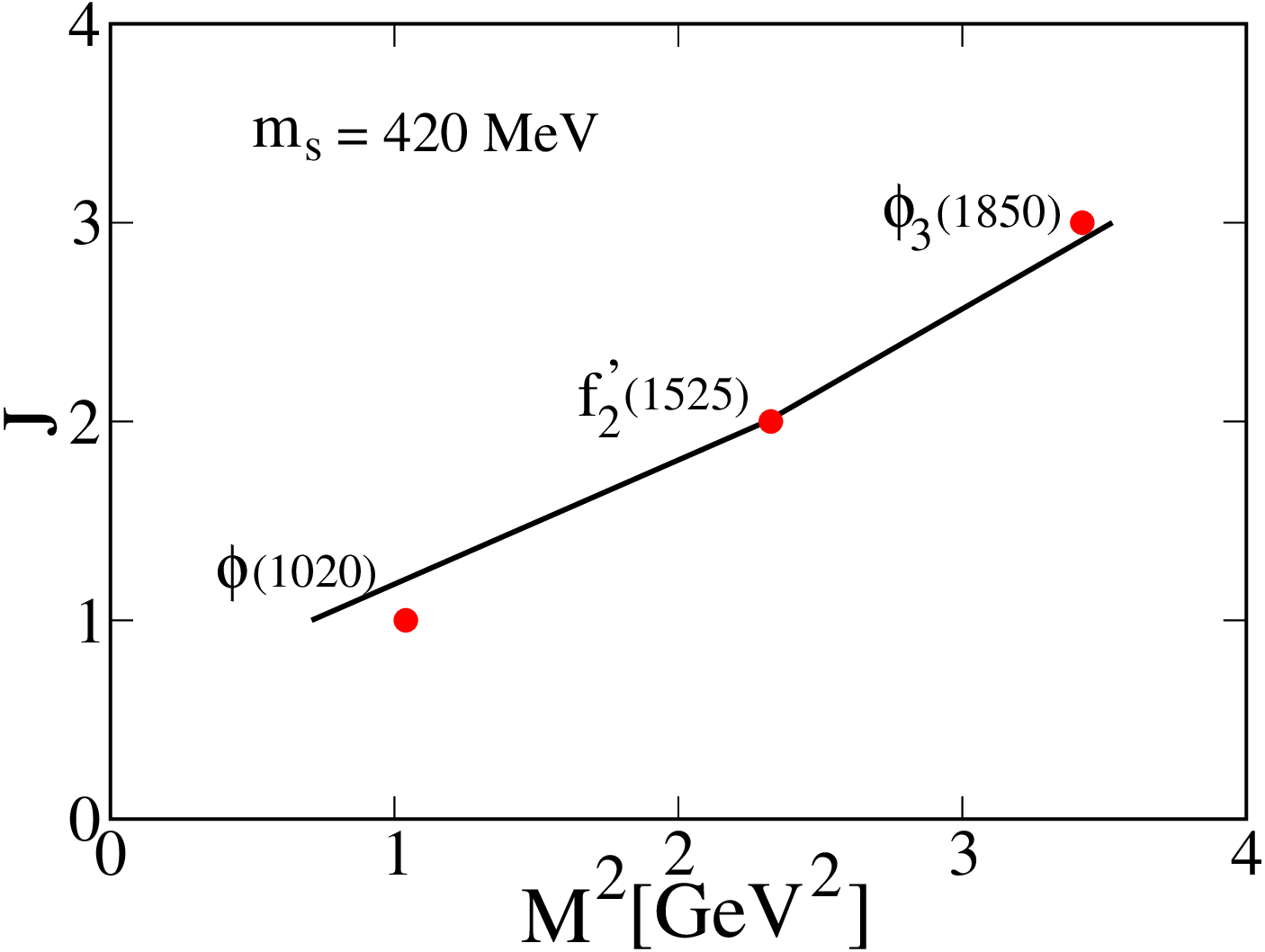}
\caption{Model predictions for the leading $\phi$ trajectory
by the relativistic
string with strange quarks at the string ends}
\label{regges}
\end{center}
\end{figure}

\begin{figure}[!htb]
\vspace{1cm}
\begin{center}
\includegraphics[width=.48\textwidth]{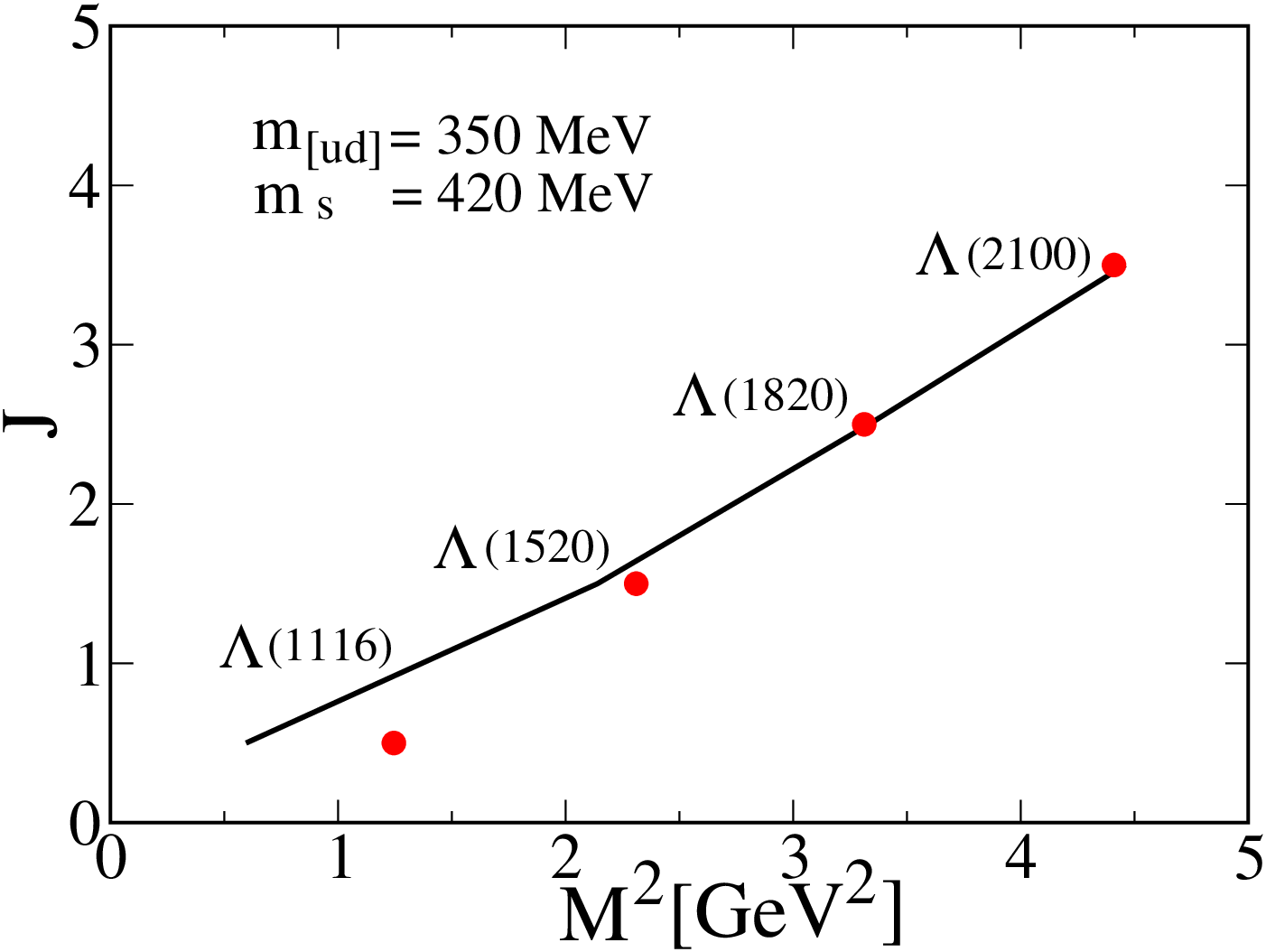}
\caption{The leading $\Lambda$ trajectory fitted by the relativistic
string with spin-isospin singlet diquark $+$ strange quark at the string ends}
\label{reggel}
\end{center}
\end{figure}

At zero orbital angular momentum ($v_i\rightarrow 0, \omega
\rightarrow \infty)$  there is no string at all and
the model mass $M$ is simply the sum of the masses of quark clusters.
As is seen from Figs. 1-4 this is not what we have experimentally
and the repulsive interaction of clusters should be included. 

Turning now to the pentaquark case and considering it as the first
orbital excitation of $[ud][ud]\bar s$ system we could imagine at least
three configurations for this excitation (see cases (A)
, (B) and (C) on Fig.5). As for the configurations (A) and (B) the masses
of clusters are here fixed from the above considerations of Regge
trajectories although in the configuration (B) the strange antiquark is
localized not at the end of the rotating string but in the centre at rest,
where its localization conditions and, hence, the constituent mass 
could be different. The masses of the configurations (A) and (B) are  determined 
by the string dynamics (in the case (B) the mass of strange antiquark is
simply added to the mass of the rotating string in eq.(\ref{string1}))
and are  equal to $1930 \rm {MeV}$ in case (A) and to
$1830 \rm {MeV}$ in case (B). By comparing these masses
 one can  conclude that configuration (A)
is unstable with respect to transformation to configuration (B).
Normally, as a result of the centrifugal forces, 
the configuration (B) should also be considered 
unstable with respect to the transformation to the configuration (C). But it is not
the case due to the strong repulsion of the antiquark $\bar s$ and 
the diquark $[ud]$. To see this let us compare the $\Lambda (1116)$ hyperon
with quark structure $[ud]s$ and the $[ud]\bar s$ cluster. From Fig.4 
(by comparing the mass of $\Lambda (1116)$ hyperon with the string result
 $m_{[ud]}+m_s$) we see that $[ud]$ and $s$ clusters repel each other
and this is not the repulsion due to spin-spin interaction which is absent
due to zero spin of $[ud]$ diquark. This repulsion could be the result of the
change of the confining conditions (change\cite{yayunyi} of the constant $B$ of 
the MIT bag model\cite{Jaffeyunyi},
for example) for $[ud]s$ system in comparison to those of $[ud]$ and $s$ clusters 
separately. Obviously, the same repulsion should be present in 
$[ud]\bar s$ cluster. If one takes the mass of $\Lambda (1116)$ hyperon
as an estimate for the mass of the $[ud]\bar s$ cluster the string dynamics
gives the mass $2080 \rm {MeV}$ for corresponding orbital excitation
(configuration (C)).

\begin{figure}[!htb]
\vspace{1cm}
\begin{center}
\includegraphics[width=.4\textwidth]{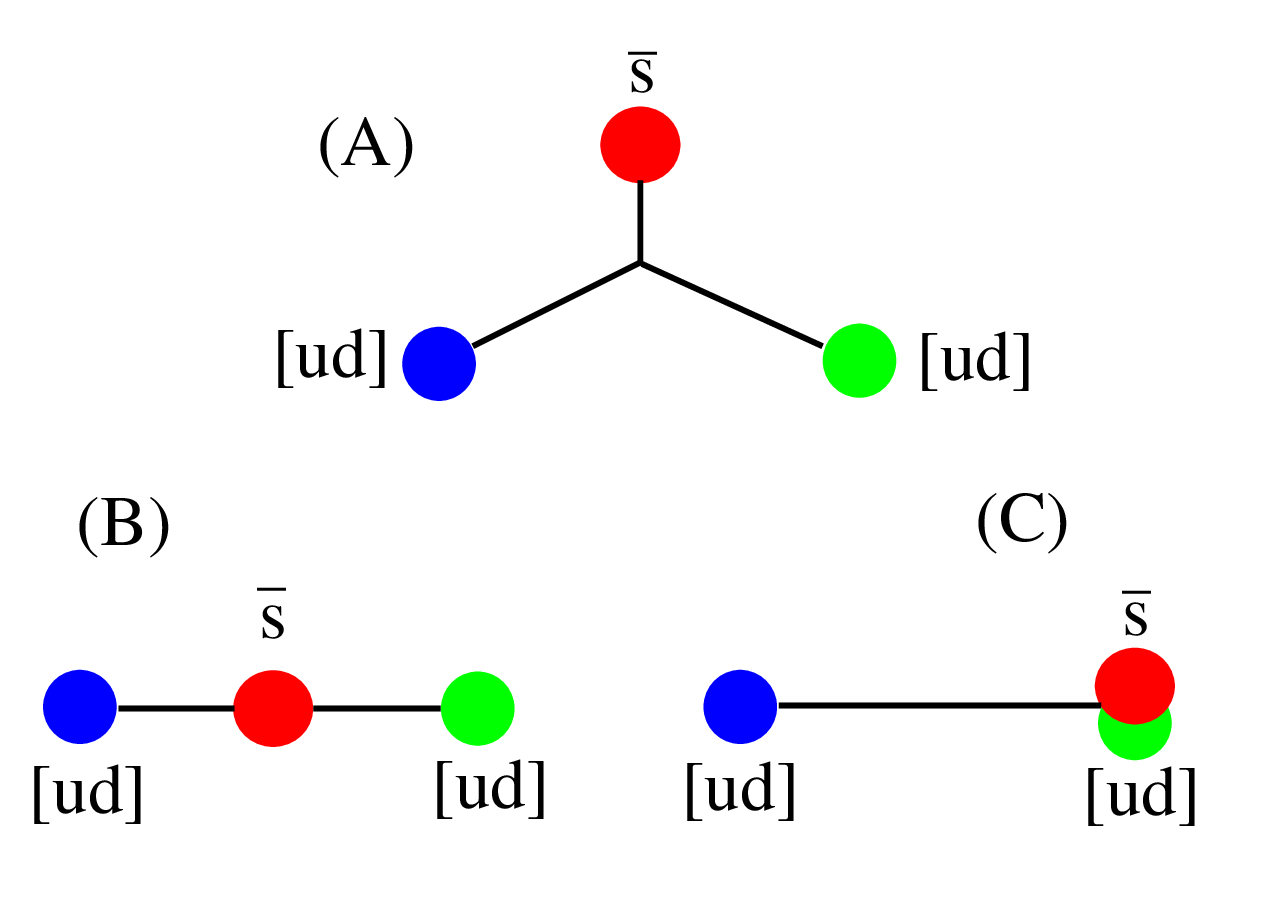}
\caption{Possible configurations of the string for
$\Theta^+(1540)$  pentaquark}
\label{reggep}
\end{center}
\end{figure}

So, we consider configuration (B) as the most reliable one for the
 pentaquark orbitally excited state. Its mass is equal
to $1830 \rm {MeV}$, $290 \rm {MeV}$ above the experimental mass.
Hence, we are forced to conclude that within the considered string model
that properly describes orbital excitations of $q\bar q$ and $qqq$ hadrons
the orbital
excitation of the pentaquark $[ud] \bar s [ud]$ is not the $\Theta^+(1540)$ particle.

One might ask how reliable is the mass $m_{[ud]} = 350 \rm {MeV}$ 
of $[ud]$ diquark extracted from $\Lambda$ hyperon trajectory.
If one tries to account for the mass difference of $\rho^+(u\bar d )$
and  $\pi^+(u\bar d )$ mesons by the spin-spin quark interaction from
gluomagnetic forces i.e. if one writes $m_{\pi} = (m_{\pi}+3m_{\rho})/4-
3(m_{\rho}-m_{\pi})/4$ and $m_{\rho} = (m_{\pi}+3m_{\rho})/4 +
(m_{\rho}-m_{\pi})/4$ ,where the first term is the mass of
the object without spin-spin interaction and the second term is
the account of spin-spin interaction,
then for the mass of spin-isospin singlet diquark we would get
$m_{[ud]} = (m_{\pi}+3m_{\rho})/4-3(m_{\rho}-m_{\pi})/8 \approx 377 \rm {MeV}$
because of the spin-spin interaction of $ud$ quarks is two times smaller
than the spin-spin interaction of $u\bar d$ quarks
due to different color functions.
This show that our estimation of the $[ud]$ diquark mass is rather reliable.

The mass of the $[ud]$ diquark that reproduces the mass of the $\Theta^+(1540)$
in the model considered is equal to $150 \rm {MeV}$. Such  a light diquark
 is not compatible with other observations (the $\Lambda$ trajectory and
 the masses of other hadrons with singlet $[ud]$ diquarks).
  Nevertheless, it is not excluded by the considerations
of dibaryons constructed from three light $[ud]$ diquarks. The reason here
is that the (spatial) wave function of three light $[ud]$ diquarks in the dibaryon
should be totally antisymmetric what means that one pair of diquarks
should have the relative orbital angular momentum equal to one ($l = 1$)
and the relative orbital angular momentum of the center of mass of the first pair
and third diquark should also be equal to one ($L = 1$). In total these
angular momenta should be summed again to one ($l+L = 1$).
This is the lowest energy state  of three diquarks 
allowed by Bose symmetry (with deuteron quantum numbers). Its totally antisymmetric
 spatial wave function has the form $(\vec {x_1}\times \vec {x_2} +
\vec {x_2}\times \vec {x_3} + \vec {x_3}\times \vec {x_1}) f(\vec {x_1},
\vec {x_2},\vec {x_3})$ where $ f(\vec {x_1},\vec {x_2},\vec {x_3})$ is the
totally symmetric under permutations of coordinates function. We don't
know which semiclassical configuration of rotating strings could correspond
to the above wave function but assume that  $l+L = 2$ configuration,
which is different from $l+L = 1$ configuration only
by the way how two units of orbital angular momenta are summed in the total
orbital angular momentum and semiclassically
corresponds  to the rotating star-like object (see Fig.5(A) with $\bar s$
substituted by $[ud]$), has
the same energy. The latter can be computed by string dynamics.
For the mass of $[ud]$ diquark needed for the reproduction of the mass
of $\Theta^+(1540)$ ($m_{[ud]} = 150 \rm {MeV}$) the mass of the 
dibaryon with deuteron quantum numbers is equal to $1910 \rm {MeV}$. It
is $30 \rm {MeV}$ above the proton-neutron threshold and the dibaryon is expected to be
as narrow as the $\Theta^+(1540)$ pentaquark. Such narrow deuteron-like dibaryon
is not observed.

In conclusion, in the simple string model for orbital excitations of light
hadrons with the masses of constituent $[ud]$ diquark and $s$ quark found
by fitting $\omega$,~$K^*$,~$\phi$,~ $\Lambda$~- trajectories
the mass of the $\Theta^+(1540)$ pentaquark, considered as a first
orbital excitation of the $[ud] \bar s [ud]$ system, is $290 \rm {MeV}$
above the experimental value. 
As have already been discussed above in the configuration (B) the strange antiquark is
localized not at the end of the rotating string but in the centre at rest,
where its localization conditions and, hence, constituent mass 
could be different.
So, in the model considered the observed discrepancy could be 
attributed to the change (the lowering) of the constituent mass of $\bar s$ antiquark as compared to
that of $s$ quark localized at the string end. 

This work is supported by RFBR grant No.
03-02-04004, DFG grant No. 436 RUS 113/721/0-1, and by Federal Program
of the Russian Ministry of Industry, Science and Technology 
No. 40.052.1.1.1112. 
M.I.K. and B.V.M.  acknowledge the kind hospitality at the University
 of Tuebingen.


\begin{thebibliography}{1}
\bibitem{1}
D. Diakonov, V. Petrov, and M. Polyakov, Z. Phys. A {\bf 359}, 305 (1997).
\bibitem{2}
LEPS Coll., T. Nakano et al., Phys. Rev. Lett. {\bf 91}, 012002 (2003).
\bibitem{3}
DIANA Coll., V. V. Barmin et al., Phys. At. Nucl. {\bf 66}, 1715 (2003).
\bibitem{4}
CLAS Coll., S. Stepanyan et al., Phys. Rev. Lett. {\bf 91}, 252001 (2003).
\bibitem{5}
CLAS Coll., V. Kubarovsky et al., Phys. Rev. Lett. {\bf 92}, 032001 (2004).
\bibitem{6}
SAPHIR Coll., J. Barth et al., Phys. Lett. B {\bf 572}, 127 (2003).
\bibitem{7}
HERMES Coll., A. Airapetian et al., hep-ex/0312044.
\bibitem{8}
SVD Coll., A. Aleev et al., hep-ex/0401024.
\bibitem{9}
COSY-TOF Coll., M. Abdel-Bary et al., hep-ex/0403011.
\bibitem{10}
A. E. Asratyan, A. G. Dolgolenko, and M. A. Kubantsev, hep-ex/0309042.
\bibitem{Jaffe}
R. Jaffe and F. Wilczek, Phys. Rev. Lett. {\bf 91}, 232003 (2003).
\bibitem{molodoiya}
I.Yu. Kobzarev, B.V. Martemyanov, M.G. Shchepkin,
Sov.Phys.Usp.{\bf 35}, 257 (1992)
\bibitem{Nambu} Y. Nambu, Phys.Rev.{\bf D10} 4262 (1974). 
\bibitem{EF} E. Eichten , F. Feinberg,
Phys.Rev. {\bf D23}, 2724 (1981).
\bibitem{yayunyi} I.Yu. Kobzarev, B.V. Martemyanov, M.G. Shchepkin,
Sov.J.Nucl.Phys. {\bf 29}, 831 (1979). 
\bibitem{Jaffeyunyi} A. Chodos, R.L. Jaffe, K. Johnson,
 Charles B. Thorn, V.F. Weisskopf, Phys.Rev.{\bf D9}, 3471 (1974).

\end{thebibliography}
 \end{document}